\documentclass[a4paper]{jpconf}
\usepackage{graphicx}
\usepackage{amssymb,amsmath,amsthm}

\newcommand{\beqn}{\begin{equation}}
\newcommand{\eeqn}{\end{equation}}

\begin{document}
\title{The hidden geometry of electromagnetism}

\author{Y Hadad$^1$, E Cohen$^2$, I Kaminer$^3$ and A C Elitzur$^4$}

\address{$^1$ Department of Mathematics, University of Arizona, Tucson, AZ, 85721, USA}
\address{$^2$ School of Physics and Astronomy, Tel Aviv University, Ramat-Aviv, Tel Aviv 6997801, Israel}
\address{$^3$ Department of Physics, Massachusetts Institute of Technology, 77 Massachusetts Avenue, Cambridge, Massachusetts 02139, USA}
\address{$^4$ Iyar, The Israeli Institute for Advanced Research, Rehovot, Israel}

\ead{yaron.hadad@gmail.com}

\begin{abstract}
Nearly all field theories suffer from singularities when particles are introduced. This is true in both classical and quantum physics. Classical field singularities result in the notorious self-force problem, where it is unknown how the dynamics of a particle change when the particle interacts with its own (self) field. Self-force is a pressing issue and an active research topic in gravitational phenomena, as well as a source of controversies in classical electromagnetism. In this work, we study a hidden geometrical structure manifested by the electromagnetic field-lines that has the potential of eliminating all singularities from classical electrodynamics. We explore preliminary results towards a consistent way of treating both self- and external fields.
\end{abstract}

%%%%%%%%%%%%%%%%%%%%%%%%%%%
\section{Introduction} \label{sec:Introduction}
The notion of a field-line was introduced by Michael Faraday in his work on magnetism \cite{bib:Faraday}. Faraday argued that the forces of electricity, magnetism and gravity are better described by fields, populated with field-lines. From Faraday's point of view, a field is not a mere mathematical abstract but a real physical object.

Today, despite Faraday's groundbreaking work, field-lines are commonly used as a mere pedagogical tool for visualizing classical fields rather than providing a genuine clue about the field's physical nature.
% \footnote{In fact, Faraday used the term `line of force'. The field line' was introduced later on.}

We suggest to revive the field's ontological status in light of several open problems in physics. Classical field theories suffer from a fundamental flaw when particles are introduced. If we naively treat the particle as point-like, the field will inevitably have singularities. For example, Maxwell's equations predict the electromagnetic field to be singular at the location of point-like electrons. Consequently, the force exerted on the electron must be infinite due to its own self-field. This is the notorious \emph{self-force problem}.

The self-force problem is ubiquitous in classical field theories. In general relativity, (gravitational) self-force is actively studied as a correction to the geodesic motion for sufficiently small bodies \cite{bib:Barack}. In classical electromagnetism, self-force introduces analogous corrections to the Lorentz force equation, one of which is the radiation-reaction force. Radiation-reaction is one of the oldest debates in physics for almost a century, with more than 15 different models suggested so far \cite{bib:Dirac}-\cite{bib:MontesCastinerias}. To date there is no compelling \emph{experimental} reason to prefer any of them. New experiments to test radiation-reaction were suggested by one of us \cite{bib:Hadad} following Labun et al., as well as other investigators \cite{bib:DiPiazza}, and will hopefully shed light on this issue in the near future.

% \cite{bib:Dirac, bib:Eliezer, bib:LandauLifshitz, bib:Prigognie-Henin1, bib:Prigognie-Henin2, bib:Nodvik, bib:Teitelboim, bib:MoPapas, bib:GonzalezGascon, bib:PetzoldSorg, bib:Caldirola, bib:FordOConnell, bib:Yaghjian, bib:Sokolov, bib:Hammond, bib:MontesCastinerias}.

% \footnote{In fact, in electrodynamics the problem of self-force is often called the problem of radiation-reaction. For the sake of precision, it is important to know that the problems differ because radiation-reaction is not the only `self-force'. For example, one should also consider the dipole moment and spin.}.

There are many ways to try and address the self-force problem in the classical regime. One is to endow the electron with an internal structure. However, such an internal structure cannot maintain its stability unless one assumes additional unknown forces \cite{bib:Rohrlich,bib:Isoyama}. A second approach is to modify Maxwell's equations to prevent the field from becoming singular in the first place. The most famous attempt of this kind is the Born-Infeld model \cite{bib:BornInfeld}, in which nonlinear effects produce an upper bound on the electromagnetic field in the vicinity of the point particle. Nevertheless, here too there is no compelling empirical reason to support the belief that the field equations are indeed nonlinear.

One may argue that the self-force problem is beyond the realm of our classical theories, hence quantum physics must be employed. In this work we present preliminary results which hint that self-forces can be treated consistently within classical physics.

Since self-force is such an elementary difficulty, it should not come as a surprise to find it strongly linked to the very concept of field-lines. By definition, a \emph{test} charge will accelerate in the direction of the tangent to the field-line that passes through it. However, pure test particles are only an idealization. In reality, every charge changes the field-lines configuration irrespectively of how small it is (see figure \ref{fig:FieldLinesExample}). This means that \emph{any (test) charge will interfere with the formation of the same field-lines it is expected to reveal}. In particular, the very concept of field-lines is ill-defined at the location of the charge, and field-lines are also inflicted by the self-force problem.

%%%%%%%%%%%%%%%%%%%%%%%%%%%%%%%%%%%%%%%%%%%%%%
\begin{figure}[h!]
\begin{center}
\includegraphics[scale=0.25]{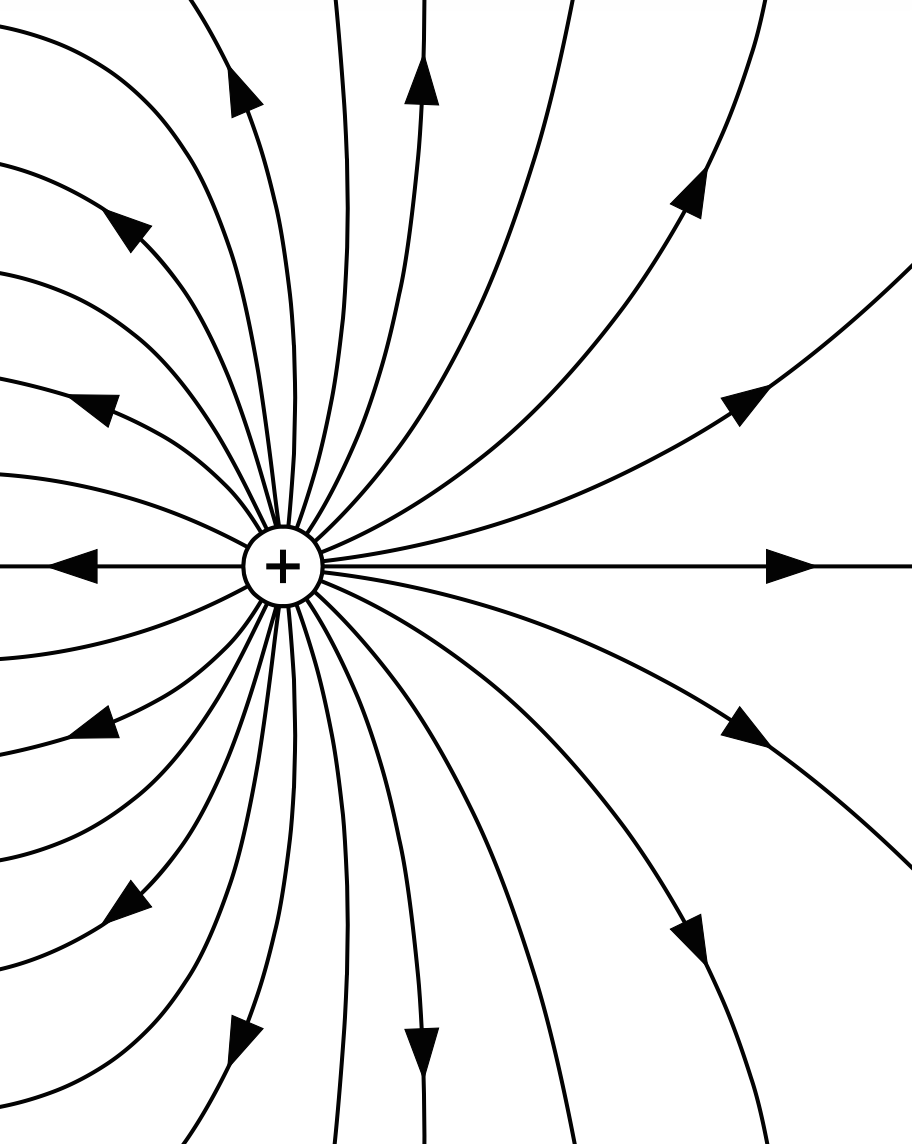}
\caption{(Image credits: BillC, Wikimedia Commons) The field-lines in the vicinity of a positive charge. By definition, a \emph{test} charge will accelerate in the direction of the tangent to the field-line at the position of the charge. The arrows point in the direction of acceleration of a positive test charge.
\label{fig:FieldLinesExample}}
\end{center}
\end{figure}
%%%%%%%%%%%%%%%%%%%%%%%%%%%%%%%%%%%%%%%%%%%%%

In the present work we argue that, paradoxically, it is the neglected notion of field-lines that offers a classical solution to the self-force problem. In earlier works, two of us \cite{bib:AvshalomEli1, bib:AvshalomEli2} adopted Faraday's viewpoint and elevated the field-lines from a mathematical abstract to a real physical entity. Just like material objects, the dynamics of the field-lines depends on their \emph{mechanical} properties. The field's dynamics is determined by the curvature of its field-lines and the stress this curvature induces. For the sake of simplicity, we proceed by studying electric field-lines in non-relativistic systems, although the ideas presented here can be generalized to other fields.

%%%%%%%%%%%%%%%%%%%%%%%%%%%
\section{Curvature and acceleration} \label{sec:Equivalence}
A free charge in (flat) empty space may be in one of two states, either at rest or traveling with a constant velocity. In the former state, the field-lines around the resting charge are isotropic. In the latter, they are denser in the directions perpendicular to the particle's velocity direction due to Lorentz contraction.

\begin{figure}[h!]
\begin{center}
\includegraphics[scale=0.24]{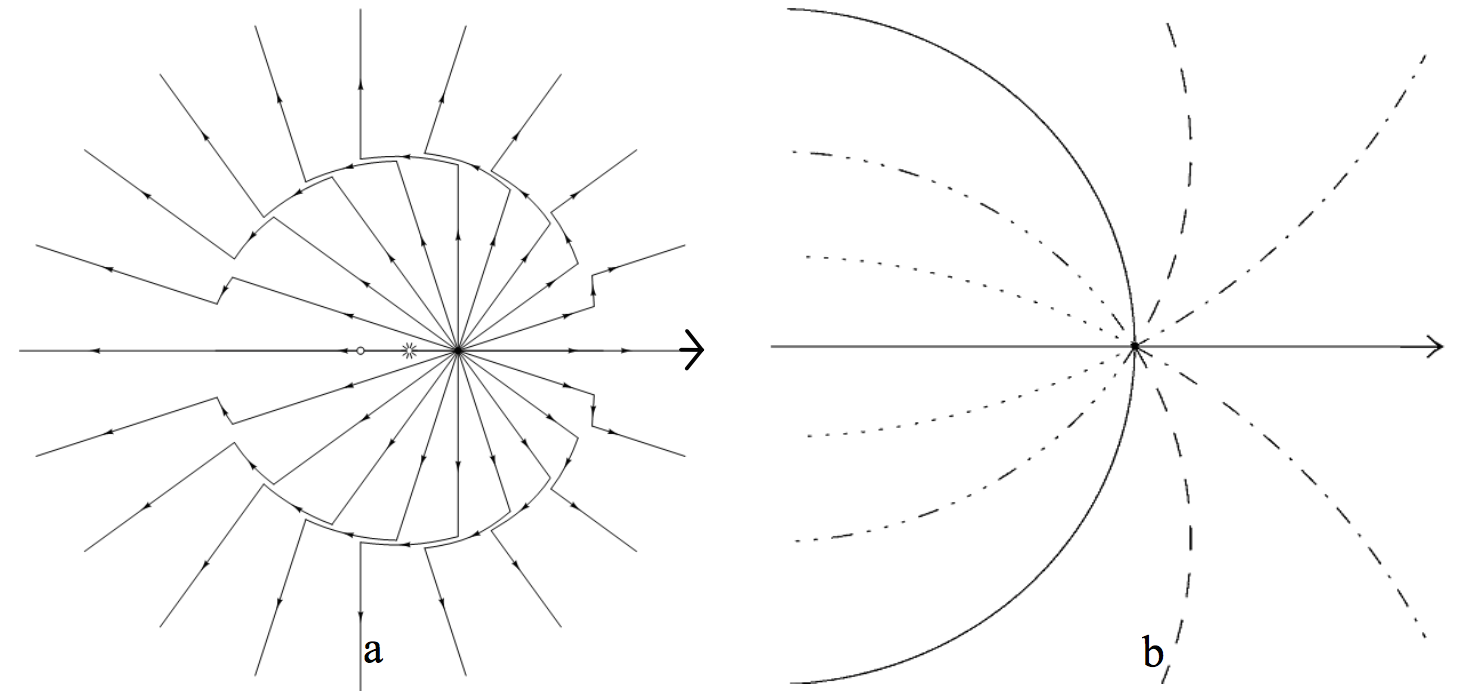}
\caption{(a) Field-lines of a charge after an instantaneous velocity change: (b) Field-lines of a charge under constant acceleration. The field-lines curve, with only those pointing in the direction of acceleration having curvature zero.
\label{fig:FieldLinesOfAcceleratingCharge}}
\end{center}
\end{figure}

Consider next a charge changing its state from the first (resting) state to the second (rectilinearly moving) state. Its velocity changes abruptly from zero to $\vec{v} = v_0 \hat{x}$, thereby emitting an electromagnetic wave. For a sufficiently far observer the charge still appears to be at rest, its field-line still being isotropic outside of the forward light-cone of the instant of acceleration. Inside the light-cone, the charge is already moving, hence the field-lines appear denser in the direction perpendicular to the direction of motion. To ensure continuity of the field-lines, they appear to be partially `broken-and-redrawn' on the light-cone's surface, as shown in figure \ref{fig:FieldLinesOfAcceleratingCharge}a.

%
% ELI I removed the reference to a step-function as it sounds contradictory. A step-function is not continuous...
%To ensure continuity of the field-lines, we see that each field line must have a step-function like behavior in the direction opposite to the direction of motion. In other words, the field lines partially breaks on the surface of the light-cone. -

This simple account can be extended to reveal a connection between the curvature of the field-lines and the charge's dynamics. In practice, the charge's velocity cannot change from zero to another value instantly but only continuously. Consider for example a velocity changing in constant rate from zero to $\vec{v} = v_0 \hat{x}$. By the same logic used above, nearly all field-lines must curve as in figure \ref{fig:FieldLinesOfAcceleratingCharge}b. The only field-lines that do not curve are the two that are directed along the charge's acceleration vector. The lines remain straight because they are on the only axis on which symmetric constraints act from both side, `pealing off' the curved lines.

%%%%%%%%%%%%%%%%%%%%%%%%%%%%%%%%%%%%%%%%%%%%%%
%%%%%%%%%%%%%%%%%%%%%%%%%%%%%%%%%%%%%%%%%%%%%

This leads to the following observations:
\begin{itemize}
\item The field-lines around an accelerating charge curve.
\item If the charge is accelerating, then the acceleration is along field-lines with zero curvature.
\end{itemize}
In the next section we will show that this observation is true not only in this specific case, but holds in general. Thus, it should be elevated to a postulate, relating the geometry of the electric field-lines to the dynamics of the charges. Under the influence of an electrostatic field, \emph{charges must travel along field-lines with zero curvature}.

%General relativity proved that formulating physical laws using geometrical ideas can be very insightful. possible that the geometrical approach described here could make meaningful predictions concerning electromagnetic phenomena?

%To address this question, we shall raise the observation previously described to the level of a postulate. The observation showed one implication (charge acceleration $\Rightarrow$ field-lines curvature), and %it is nothing but tempting to hope that an equivalence between the two statements hold. Thus, we postulate:

%{{{
%\emph{``The maximal acceleration of a charge is equivalent to the curvature of the field lines in the neighborhood of the charge.''}
%}}}
%We will refer to this postulate as the \emph{curvature-acceleration equivalence}. The postulate is both vague and probably wrong...
%(NOTE: I DON'T LIKE THE LAST TWO PARAGRAPHS)

%%%%%%%%%%%%%%%%%%%%%%%%%%%
\section{Minimal (zero) curvature} \label{sec:ZeroCurvature}
We start by considering a physical system which includes charges traveling under the influence of an electric field. For the sake of simplicity, we assume no magnetic fields or other forces. Our goal is to study the geometrical properties of the electric field-lines and their relation to the dynamics of the charges.

An electric field-line is defined to be a curve in $\mathbb{R} ^3$ for which the direction of the curve is equal to the direction of the electric field. In other words, if $\vec{\gamma}(t, s)$ is an electric field-line at time $t$ with parametrization $s$, then $\vec{\gamma}$ must satisfy the \emph{electric field-line equation}:
\begin{equation} \label{eq:ElectricFieldLine}
\frac{d}{ds} \vec{\gamma}(t, s) = \vec{E} (t, \vec{\gamma}(s)).
\end{equation}
The curvature $\kappa$ of the curve $\vec{\gamma}$ in $\mathbb{R} ^3$ is given by a well-known formula from differential geometry:
\begin{equation} \label{eq:CurvatureOfCurve}
\kappa = \frac{|\vec{\gamma}'\times \vec{\gamma}''|}{|\vec{\gamma}'|^3},
\end{equation}
where the prime denotes differentiation with respect to the parametrization $s$. It should be clear that both functions $\kappa$ and $\vec{\gamma}$ depend on the time $t$ and the parametrization $s$, but to keep the notation concise we will omit these arguments.

By replacing the electric field-line Eq. (\ref{eq:ElectricFieldLine}) in the curvature Eq. (\ref{eq:CurvatureOfCurve}), we receive the curvature $\kappa(t,\vec{x})$ of the unique field-line that passes through $\vec{x}$ at time $t$,
\begin{equation}
\kappa(t, \vec{x}) = \frac{|\vec{E} \times (\vec{E} \cdot \nabla) \vec{E}|}{|\vec{E}|^3}
\end{equation}
where the electric field $\vec{E}$ is sampled at the point $\vec{x}$ and time $t$.

In order to assess the validity of our observation of the previous section we need to evaluate the curvature near the position $\vec{x_0} (t)$ of the charge. In classical electromagnetism it is common to omit the self-fields of the charges, to avoid the self-force problem that leads to the notorious singularities as well as the radiation-reaction problem \cite{bib:Rohrlich}. But here the situation is fundamentally different. \emph{When dealing with the geometry of the field-lines, one must include the self-fields of the charge to obtain meaningful results}. In other words, we must treat the electric field as a whole.

Let us separate the electric field into two parts. The first is the external electric field $\vec{E} _{\text{ext}}$. Here `external' refers to the contribution of the electric field from all sources other than the charge located at the point $\vec{x_0}(t)$. The self-field of this charge will be denoted by $\vec{E}_{\text{self}}$. The total electric field is the superposition,
\begin{equation} \label{eq:TotalField}
\vec{E} = \vec{E}_{\text{self}} + \vec{E}_{\text{ext}}.
\end{equation}
Substituting Eq. (\ref{eq:TotalField}) in the curvature Eq. (\ref{eq:CurvatureOfCurve}) yields
\begin{equation}
\kappa(t, \vec{x}) = \frac{|(\vec{E}_{\text{self}} + \vec{E}_{\text{ext}}) \times ((\vec{E}_{\text{self}} + \vec{E}_{\text{ext}}) \cdot \nabla) (\vec{E}_{\text{self}} + \vec{E}_{\text{ext}})|}{|\vec{E}_{\text{self}} + \vec{E}_{\text{ext}}|^3}.
\end{equation}

Coulomb law shows that in the immediate vicinity of the charge the self-field is greater in magnitude than the external field, while farther away the external field gives a greater contribution. This means that in the nearby neighborhood of the charge, the dominating factor in the last equation is $\vec{E}_{\text{self}}$. We may therefore expand the curvature $\kappa$ about the charge's position $\vec{x}_0 (t)$ by treating the self-field as a dominating term.

In the zeroth order, the curvature is determined by the self-field alone, and we have
\begin{equation}
\kappa(t, \vec{x}) \approx \frac{|\vec{E}_{\text{self}} \times (\vec{E}_{\text{self}} \cdot \nabla) \vec{E}_{\text{self}}|}{|\vec{E}_{\text{self}}|^3}.
\end{equation}
By Coulomb's law the self field-lines are straight. Hence this term vanishes for all $\vec{x} \neq \vec{x}_0 (t)$, and one must consider the next order in the perturbation expansion.

The next order in the perturbation gives the first non-zero contribution to the curvature and is given by
\begin{equation}
\kappa(t, \vec{x}) = \frac{|\vec{k}_1 + \vec{k}_2 + \vec{k}_3|}{|\vec{E}_{\text{self}}|^3},
\end{equation}
where
\begin{equation} \label{eq:k1}
\vec{k}_1 = \vec{E}_{\text{ext}} \times (\vec{E}_{\text{self}} \cdot \nabla) \vec{E}_{\text{self}},
\end{equation}
\begin{equation} \label{eq:k2}
\vec{k}_2 = \vec{E}_{\text{self}} \times (\vec{E}_{\text{ext}} \cdot \nabla) \vec{E}_{\text{self}},
\end{equation}
\begin{equation} \label{eq:k3}
\vec{k}_3 = \vec{E}_{\text{self}} \times (\vec{E}_{\text{self}} \cdot \nabla) \vec{E}_{\text{ext}}.
\end{equation}
Note that each of the terms $\vec{k}_1$, $\vec{k}_2$ and $\vec{k}_3$ has two appearances of the self-field and a single appearance of the external field.

Let us assume that we are in the instantaneous reference frame of the charge. In this reference frame Coulomb's law holds and the self-field is
\begin{equation}
\vec{E} _{\text{self}} (t,\vec{x}) = q \frac{\vec{x}_0 (t) - \vec{x}}{|\vec{x}_0 (t) - \vec{x})|^3}.
\end{equation}

Plugging this result into Eqs. (\ref{eq:k1}, \ref{eq:k2}) gives
\begin{equation}
\vec{k}_1 (t,\vec{x}) = \frac{2q^2}{|\vec{x}_0 (t) - \vec{x}|} \vec{E}_{\text{ext}} (t,\vec{x}) \times (\vec{x}_0 (t) - \vec{x}),
\end{equation}
\begin{equation}
\vec{k}_2 (t,\vec{x}) = \frac{q^2}{|\vec{x}_0 (t) - \vec{x}|} \vec{E}_{\text{ext}} (t,\vec{x}) \times (\vec{x}_0 (t) - \vec{x}).
\end{equation}
Similarly, Eq. (\ref{eq:k3}) shows that $\vec{k}_3$ is of a smaller magnitude ($1/r^4$) than $\vec{k}_1$ and $\vec{k}_2$ ($1/r^5$) due to the appearance of the derivative of external field in Eq. (\ref{eq:k3}). $\vec{k}_3$ may therefore be omitted.

The above calculation shows that
\begin{equation} \label{eq:ElectricCurvature}
\kappa(t, \vec{x}) \approx \frac{3}{q} |\vec{E} _{\text{ext}}(t, \vec{x}) \times (\vec{x}_0 (t) - \vec{x})|
\end{equation}
about the charge and in its rest frame. Remarkably, we see that any direct dependence on the self-field of the charge actually vanished. In the neighborhood of the charge, the curvature of the electric field-lines depends on the external field (as well as the vector $\Delta \vec{x} = \vec{x}_0 (t) - \vec{x}$). This has profound consequences, as Eq. (\ref{eq:ElectricCurvature}) shows that the electric field-lines curvature is regular nearby the charge. This result holds for all electric charges, and therefore shows that the electric field-lines curvature is always regular. In other words, even \emph{for point-like charged particles the electric curvature is a well-defined property}.

%%%%%%%%%%%%%%%%%%%%%%%%%%%%%%%%%%%%%%%%%%%%%%
\begin{figure}[h!]
\begin{center}
\includegraphics[scale=0.17]{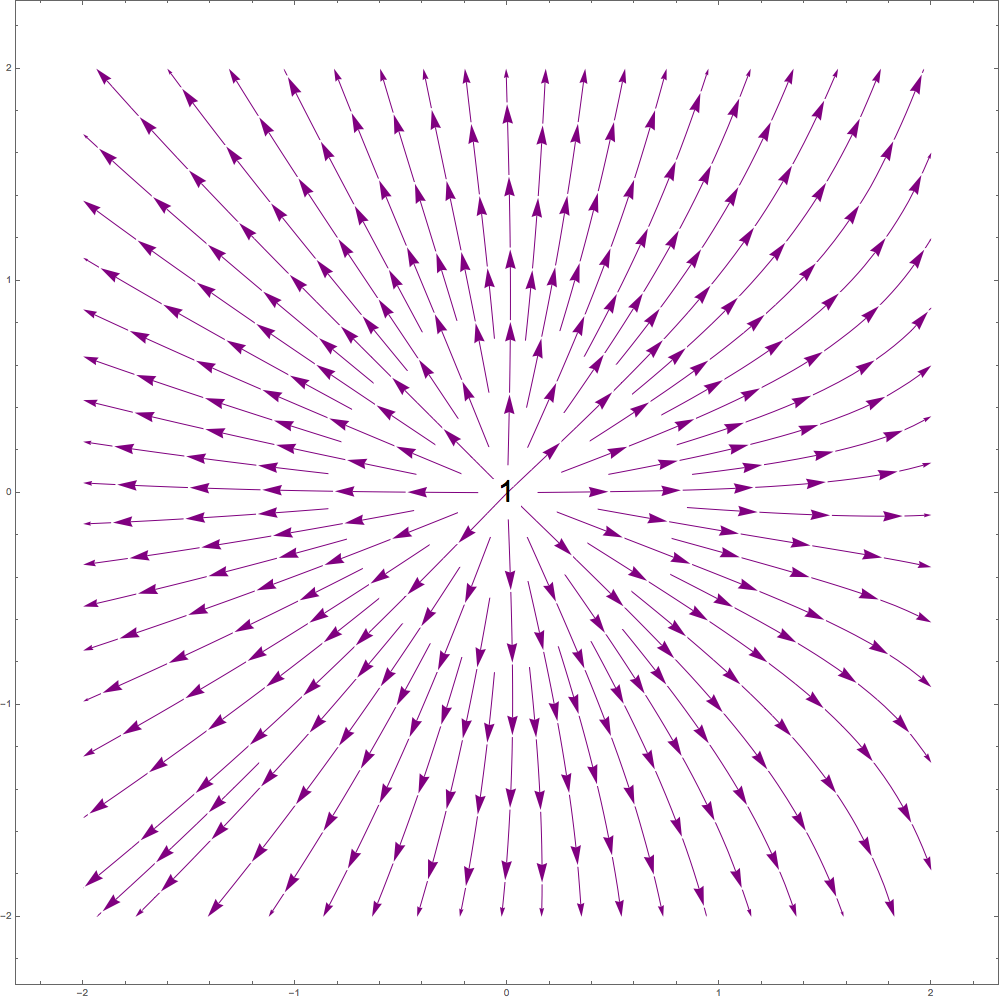}
\includegraphics[scale=0.17]{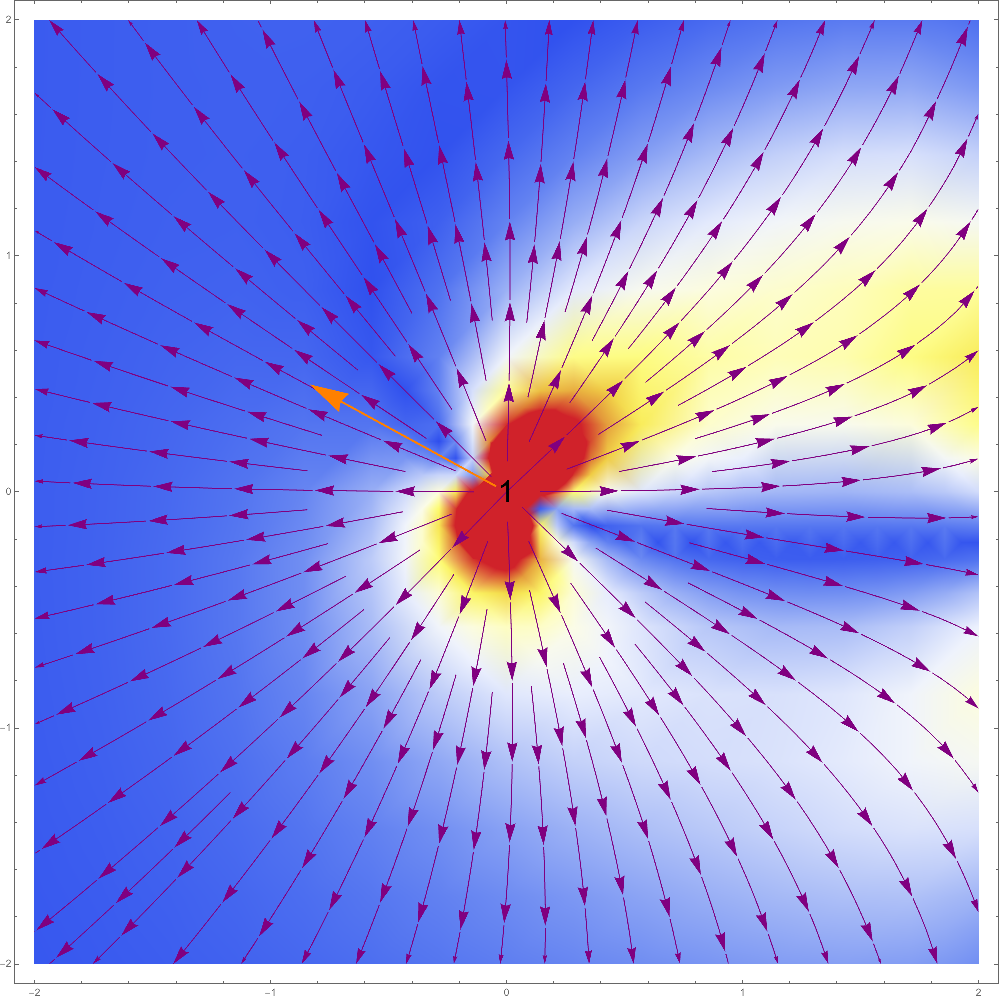}
\includegraphics[scale=0.6]{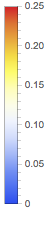}
\caption{(a) The field-lines around a positive charge. In the immediate vicinity of the charge the field-lines are mostly determined by its self-field, where farther away it is the external field that predominantly determines the field-lines. (b) Each field-line is a curve in $\mathbb{R}^3$, and in this figure the color of each point represents the value of the field-lines' curvature. The blue color indicates regions where the field-line curvature vanishes, and it is clear that the acceleration vector of the charge (the orange arrow) points in the direction of zero curvature.
\label{fig:FieldLinesExample}}
\end{center}
\end{figure}
%%%%%%%%%%%%%%%%%%%%%%%%%%%%%%%%%%%%%%%%%%%%%

Eq. (\ref{eq:ElectricCurvature}) provides several immediate results. First, the limiting case $q\rightarrow \infty$ gives $\kappa\rightarrow 0$. This is not surprising, as in this case the self-field dominates the motion and the field-lines are straight.

The second limiting case $q\rightarrow 0$ is singular for this formula. In this case Eq. (\ref{eq:ElectricCurvature}) is no longer valid as the assumptions we used in its derivation no longer hold.
This is because we assumed $q\neq 0$ when we started this perturbation expansion, and for this case the equation is no longer valid since the external field dominates over the self-field, and the relative orders of terms $k_1$,$k_2$,$k_3$ changes.  Furthermore, the relative magnitudes of $E_{ext}$ and $q$ provides a natural length dimension (including the curvature radius) which separates between the `near' and the `far' zones.

In the last section we observed that a charge accelerates in the direction of field-lines with zero curvature. We are now ready to prove that this is a general phenomenon. Let $\Delta t$ be an infinitesimal time duration. Consider evaluating the curvature of the field at the next time step of the charge,
\begin{equation}
\vec{x} = \vec{x}_0 (t + \Delta t).
\end{equation}
For small $\Delta t$ we have,
\begin{equation} \label{eq:Dx}
\vec{x}_0 (t + \Delta t) - \vec{x}_0 (t) \approx \vec{v}_0 (t) \Delta t + \frac{1}{2} \vec{a}_0 (t) (\Delta t)^2.
\end{equation}

The curvature formula (\ref{eq:CurvatureOfCurve}) was derived in the instantaneous rest frame of the charge, in which $\vec{v}_0 = \vec{0}$. The curvature (\ref{eq:CurvatureOfCurve}) at the position of the charge at time $t + \Delta t$ can be estimated using Eq. (\ref{eq:Dx}) (where $\vec{v}_0 (t) = \vec{0}$) to give
\begin{eqnarray}
\kappa(t, \vec{x}_0 (t+ \Delta t)) &\approx& \frac{3}{q} |\vec{E} _{\text{ext}}(t, \vec{x}) \times (\vec{x}_0 (t) - \vec{x}_0 (t+ \Delta t))| \\ \notag
&\approx& \frac{3 (\Delta t)^2}{2q} |\vec{E} _{\text{ext}}(t, \vec{x_0} (t + \Delta t)) \times \vec{a}_0 (t)|.
\end{eqnarray}
This means that $\kappa(t, \vec{x}_0 (t+ \Delta t)) = 0$ if and only if the acceleration is parallel to the external field. \emph{The charge will always accelerate in the direction of a field-line that has zero curvature}.

The most striking result from the curvature Eq. (\ref{eq:CurvatureOfCurve}) is that the electric curvature is regular nearby point charges. This is a surprising result, as most electromagnetic quantities are singular near point particles \cite{bib:Dyson}. This preliminary result hints on a consistent way of dealing with point particles without the renormalization techniques that are typically employed in the literature \cite{bib:Peskin}.

%%%%%%%%%%%%%%%%%%%%%%%%%%%%%%%%%%%%%%%%%%
\section{Maximal curvature} \label{sec:MaximalCurvature}

The acceleration of the charge is a vector. By its definition, the acceleration is uniquely determined by its direction and magnitude. In the last section we saw that the geometrical properties (i.e. curvature) of the field-lines suffice to determine the direction of the acceleration. The goal of this section is to demonstrate that the field-lines also determine the acceleration's magnitude.

Consider figure \ref{fig:FieldLinesOfAcceleratingCharge}a again. The greater the difference between the initial and final velocities, the more broken the field-lines will appear on the light cone's surface. This implies that there is a relationship between the magnitude of the charge's acceleration and the curvature of the field-lines in the vicinity of the accelerating charge. In other words, as the acceleration becomes higher the field-lines nearby the charge are more curved (see \ref{fig:FieldLinesOfAcceleratingCharge}b).

There are several approaches to measure the total curvature of the field-lines about the charge. Here we present only one, not necessarily the most natural nor the most general. It is considered only as a case study that shows that \emph{the field-lines contain all the information needed to know the dynamics of the charge}.
% The most natural approach is typically an action principle, which will be the last method considered here.

In the last section, the field-lines curvature $\kappa$ in the neighborhood of the charge was shown to determine the direction of the charge's acceleration vector. This was done by requiring that the charge accelerates along a vector of \emph{zero electromagnetic curvature}. It is clear that lines of zero curvature cannot determine the magnitude of the acceleration vector. Since the electromagnetic curvature is always non-negative, the zero curvature condition isolates the \emph{minimal} curvature in the neighborhood of the charge. A natural contra-distinctive way of determining the magnitude of the acceleration is to use the \emph{maximum} curvature in the neighborhood of the charge.

The curvature of the electromagnetic field at the point $\vec{x}$ was computed in Eq. (\ref{eq:ElectricCurvature}), and can be rewritten as
\begin{equation} \label{eq:ElectricCurvature2}
\kappa(t, \vec{x}) \approx \frac{3}{q} |\vec{E} _{\text{ext}}(t, \vec{x})| |(\vec{x}_0 (t) - \vec{x})| |\sin \theta|,
\end{equation}
where $\theta$ is the angle between the electric field-line at the point $\vec{x}$ and the vector $\vec{x}_0 (t) - \vec{x}$. Assume that the radius of the ball is small enough so that the electric field barely varies inside the ball,
\begin{equation}
\vec{E}_{\text{ext}}(t, \vec{x}) \approx \vec{E}_{\text{ext}}(t, \vec{x}_0).
\end{equation}
Under this assumption, Eq. (\ref{eq:ElectricCurvature2}) shows that the maximal curvature in a ball of radius $r$ about $x_0 (t)$ is obtained along the angle $\theta=\pi/2$, corresponding to a perpendicular direction to the acceleration vector and is given by
\begin{equation} \label{eq:MaximumCurvature}
\max_{|\vec{x}-\vec{x}_0(t)| \leq r} \,\kappa = \frac{3}{q} |\vec{E} _{\text{ext}}(t, \vec{x}_0)| r.
\end{equation}
Eq. (\ref{eq:MaximumCurvature}) shows that the maximal curvature is proportional to the magnitude of the total electric field. It is well-known that in the absence of other forces, the total electric field is proportional to the magnitude of the acceleration of the charge.

We therefore postulate that \emph{the charge accelerates in the direction of zero curvature field-lines with acceleration magnitude proportional to the maximal curvature of the field (line)}. The proportionality constant can be determined using dimensional analysis. The dimensions of curvature are one over length. To obtain dimensions of acceleration we need to multiply the curvature $\kappa$ by a constant with dimensions of velocity squared, and it is only natural to multiply it by $c^2$ (here $c$ denotes the speed of light propagation in vacuum).

The only variable left to determine is the radius $r$ used in Eq. (\ref{eq:MaximumCurvature}). Any result produced by the method suggested here must be consistent with the Lorentz force in the absence of magnetic fields,
\begin{equation} \label{eq:LorentzForce}
m \vec{a} = q \vec{E}.
\end{equation}
It is now easy to see that setting
\begin{equation} \label{eq:r}
r = \frac{1}{3} \frac{q^2}{m c^2}
\end{equation}
in Eq. (\ref{eq:MaximumCurvature}) produces the expected results (\ref{eq:LorentzForce}) in the leading order. Interestingly, for electrons the radius we obtained in Eq. (\ref{eq:r}) is
\begin{equation}
r = \frac{1}{3} r_e,
\end{equation}
where
\begin{equation}
r_e = \frac{e^2}{m_e c^2} = 2.8179403267(27)\times 10^{-13} \,\text{cm}
\end{equation}
is the classical electron radius. Here $e$ and $m_e$ are the electric charge and mass of the electron respectively.

%%%%%%%%%%%%%%%%%%%%%%%%%%%%%%%%%
\section{Conclusions} \label{sec:Conclusions}

The notion of field-lines is widely discussed in textbooks as a tool for visualizing fields. In this work, we argued to the contrast. Rather than a mere visualization tool, field-lines may causally determine the dynamics of moving particles. This postulate was proven for electric fields, as it was shown that the electric field-lines completely determine the dynamics of electric charges. The curvature of the field-lines as curves in $\mathbb{R}^3$ contains all the information needed to determine the acceleration of electric charges under an electrostatic force.

We studied an electric charge $q$ moving under the influence of both its self-field $\vec{E} _\text{self}$ and an external electric field $\vec{E} _\text{ext}$. The curvature of the field-lines was always computed by considering the total electric field $\vec{E} = \vec{E}_{\text{self}} + \vec{E}_{\text{ext}}$. The acceleration of the charge was shown to be directed along the minimum (zero) field-line curvature, and its magnitude is proportional to the maximum field-line curvature. Written explicitly,
\begin{equation}
\hat{a} = \text{direction of } \min( \kappa) = \hat{E},
\end{equation}
and
\begin{equation}
|\vec{a}| = c^2 \max ( \kappa) = \frac{q}{m} |\vec{E} _\text{ext} (t, \vec{x}_0)|
\end{equation}
where $c$ is the speed of light, $m$ is the mass of the charge, $\vec{x}_0$ is its position, and both the minimum and maximum curvature are computed within a ball of a radius given in Eq. (\ref{eq:r}). For electrons, this radius is equal to one third of the classical electron radius.

The most surprising result from this analysis is that although we studied the curvature of the total electric field $\vec{E}$, the self-field produced no singularities in the total curvature. This is apparent from Eq. (\ref{eq:ElectricCurvature}) which only depends on the external field $\vec{E}_\text{ext}$. This result holds independently of the electric charge's structure, and continues holding for point charges. This hints that the formalism studied here is capable of treating point particles consistently without the need for renormalization. Since field singularities constitute an inherent issue when studying field theories that contain particles, this analysis appears to be a promising research direction for eliminating singularities from field theories in general.

The results provided here are still preliminary, and many interesting questions arise:
\begin{itemize}
\item How can the above mechanism be extended to treat fields other than the electric field? The most natural next step would be to study the magnetic field. Should that prove successful, venturing straight ahead to gravity would be the next natural step.
\item How can these results be derived in Minkowski spacetime using the 4-vector definition of curvature on general electromagnetic fields? The results produced here are indeed consistent with the special theory of relativity, and a robust relativistic formulation of them can be very insightful.
\item Can the field-lines be used to obtain an action principle for classical electrodynamics that treats the self-fields on the same footing as the external fields without apparent singularities?
\item What about quantum fields?
\end{itemize}

It is important to remember that in this work we approximated the maximal curvature in Eq. (\ref{eq:MaximumCurvature}) in order to derive the Lorentz force Eq. (\ref{eq:LorentzForce}). It is only expected that the full non-approximated expression will include correction terms. We hope that these correction terms will shed light on the problem of radiation-reaction, which remains an open problem in classical electrodynamics.

\hfill

\ack{We thank Ofek Birnholtz, Paz Beniamini, Michael Bialy, Doron Grossman and Larry Horwitz for very helpful comments and discussions. E.C. was supported in part by Israel Science Foundation Grant No. 1311/14. I.K. was supported by a Marie Curie Grant No. 328853-MC-BSiCS.}

\medskip

\hfill
\hfill


\begin{thebibliography}{42}

\bibitem{bib:Faraday}
Faraday M 1846 {\it Philos. Mag.} {\bf 28} 345

\bibitem{bib:Barack} Barack L 2014 {\it Fund. Theor.} {\bf 177} 147

\bibitem{bib:Dirac}
Dirac P A M 1938 {\it Proc. R. Soc. Lon. Ser. A} {\bf 167} 148

\bibitem{bib:Eliezer}
Eliezer C 1948, {\it Proc. R. Soc. Lon. Ser. A} {\bf 194} 543

\bibitem{bib:LandauLifshitz}
Landau L D and Lifshitz E M 1975 {\it The Classical Theory of Fields} (Oxford: Elsevier)

\bibitem{bib:Prigognie-Henin1}
Prigogine I and Henin F 1962 {\it Physica} {\bf 28} 667

\bibitem{bib:Prigognie-Henin2}
Prigogine I  and Henin F 1963 {\it Physica} {\bf 29} 286

\bibitem{bib:Nodvik}
Nodvik J S 1964 {\it Ann. Phys.} {\bf 28} 225

\bibitem{bib:Teitelboim}
Teitelboim C 1971 {\it Phys. Rev. D} {\bf 4} 45

\bibitem{bib:MoPapas}
Mo T C and Papas C H 1971 {\it Phys. Rev. D} {\bf 4} 3566

\bibitem{bib:GonzalezGascon}
Gonzalez-Gascon F 1976 {\it Nuovo Ciemnt. B} {\bf 11} 333

\bibitem{bib:PetzoldSorg}
Petzold J and Sorg M 1977 {\it Z. Phys. A} {\bf 283} 207

\bibitem{bib:Caldirola}
Caldirola A 1979 {\it Riv. Nuovo Cimento} {\bf  2} 1

\bibitem{bib:FordOConnell}
Ford G W and R F O'Connell 1991 {\it Phys. Lett. A} {\bf 157} 217

\bibitem{bib:Yaghjian}
Yaghjian A D 1992 {\it Relativistic Dynamics of a Charged Sphere} (Berlin: Springer-Verlag)

\bibitem{bib:Sokolov}
Sokolov I V, Naumova N M, Nees J A, Mourou G A and Yanovsky V P 2009 {\it Phys. Plasmas} {\bf 16} 093115

\bibitem{bib:Hammond}
Hammond  R T 2010 {\it Elect. J. Theor. Phys.} {\bf 6} 221

\bibitem{bib:MontesCastinerias}
de Oca A C M and Castineiras J 2013 On radiation reaction and the Abraham-Lorentz-Dirac equation {\it Preprint} arXiv:1304.2203

%\bibitem{bib:Ofek}
%Birnholtz O 2015  {\it Int. J. Mod. Phys. A}  {\bf 30} 1550011

\bibitem{bib:Hadad}
Hadad Y, Labun L Rafelski J, Elkina N, Klier C and Ruhl H 2010 {\it Phys. Rev. D} {\bf 82} 096012

\bibitem{bib:DiPiazza}
Di Piazza A, M$\ddot{u}$ller C, Hatsagortsyan K Z and Keitel C H 2012 {\it Rev. Mod. Phys.} {\bf 84} 1177

\bibitem{bib:Rohrlich}
Rohrlich F 1997 {\it Am. J. Phys.} {\bf 65} 1051

\bibitem{bib:Isoyama}
Isoyama S and Poisson E 2012 {\it Classical Quant. Grav.} {\bf 29} 155012

\bibitem{bib:BornInfeld}
Born M and Infeld L 1934 {\it Proc. R. Soc. Lon. Ser. A} {\bf 144} 425

%\bibitem{bib:Greiner}
%Greiner W 1985 {\it Springer Berlin Heidelberg}

\bibitem{bib:AvshalomEli1}
Elitzur A C, Cohen E and Beniamini P 2012 Charge Acceleration and Field-Lines Curvature: A Fundamental Symmetry and Consequent Asymmetries {\it AIP Conf. Proc.} {\bf 1411} 211 ({\it Preprint} arXiv:1208.5164)

\bibitem{bib:AvshalomEli2}
Cohen E, Beniamini P, Grossman D, Horwitz L and Elitzur A C 2013 Mechanical Properties of the Electric Field: A Novel Prediction derived from the Field's Mass and Stress {\it Preprint} arXiv:1304.5598

\bibitem{bib:Dyson}
Dyson F J 1952 {\it Phys. Rev.} {\bf 85} 631

\bibitem{bib:Peskin}
Peskin M E and Schroeder D V 1995 {\it An introduction to quantum field theory} (Boulder: Westview)

%\bibitem{Poisson} Poisson. An introduction to the Lorentz-Dirac equation. Arxiv preprint gr-qc (1999)
%\bibitem{RohrlichNew} Rohrlich. Dynamics of a charged particle. Physical Review E (2008)

\end{thebibliography}
\end{document}